\documentclass[aps,pra,reprint,groupedaddress,showpacs]{revtex4-1}%
\usepackage{amsmath}
\usepackage{graphicx}
\usepackage{textcomp}
\usepackage{hyperref}
\usepackage{amsfonts}
\usepackage{amssymb}%
\providecommand{\U}[1]{\protect\rule{.1in}{.1in}}
\pdfoutput=1
\begin{document}
\title{Reactive energy in non-diffracting localized waves}
\author{Peeter Saari$^{1,2}$}
\email{Corresponding author: peeter.saari@ut.ee}
\author{Ioannis Besieris$^{3}$}
\affiliation{$^{1}$Institute of Physics, University of Tartu, W. Ostwaldi 1, 50411, Tartu, Estonia}
\affiliation{$^{2}$Estonian Academy of Sciences, Kohtu 6, 10130 Tallinn, Estonia}
\affiliation{$^{3}$The Bradley Department of Electrical and Computer Engineering, Virginia
Polytechnic Institute and State University, Blacksburg, Virginia 24060, USA}
\date{\today}

\begin{abstract}
It is well known that although the group velocity of structured light pulses
propagating in vacuum can be subluminal or superluminal, the upper limit of
the energy flow velocity is $c$, the speed of light in vacuum. This inequality
can be explained in terms of the reactive energy left behind by the fields.
Energy and reactive energy densities have been calculated for vector-valued
two-dimensional (light sheet) superluminal electromagnetic nondiffractive
pulses, as well as scalar-valued and TM three-dimensional superluminal and
subluminal spatiotemporally localized electromagnetic waves. Emphasis is
placed on the physical formation of the reactive energy due to interference of
the plane-wave constituents of the structured light waves.

\end{abstract}

\pacs{42.25.Bs, 42.25.Fx, 42.60.Jf, 42.65.Re}
\keywords{Bessel beam; Bessel-X pulse; group velocity; energy velocity; Poynting vector}\maketitle

\section{Introduction}

It is well known that in the case of electromagnetic (EM) one-dimensional
plane wave pulses in vacuum, not only the phase velocity but also the group
and energy flow velocities are equal to the universal constant $c$. But it is
not so when the wave is spatially confined, i.e., generally in the case of
structured light. For example, in a pulsed Gaussian beam, the group velocity
$v_{g}$ may not be only luminal $(v_{g}/c=1),$ but also slightly superluminal
$(v_{g}/c>1)$ or subluminal $(v_{g}/c<1)$ locally in the beam waist region.
Which alternative takes place depends on how the frequency and the waist width
of every monochromatic Gaussian beam constituent of the pulsed beam are
coupled with each other. Subtleties of relations between the pulse group
velocity and the measurable travel time, as well as some relevant
contradictions in the recent literature, are analyzed in
Ref.~\cite{MinuPRA2018}.

A wide class of structured light pulses, called non-diffracting localized
waves, is characterized by a specific type of space-time coupling. For all
monochromatic plane-wave constituents of such pulsed waves, there is a linear
functional dependence between their temporal frequency $\omega$ and the
component $k_{z}$ of the wave vector, which lies in the direction of
propagation of the pulses, rendering the latter propagation-invariant. This
means that the spatial distribution of the pulse energy density does not
change in the course of propagation---it does not spread either in the lateral
or in the longitudinal direction (or temporally). In reality, such a
non-diffracting (non-spreading) propagation occurs over a large but still
finite distance, because the aforementioned frequency-wavenumber functional
dependence is not strict for practically realizable (finite-energy and finite
aperture) pulses.

The first versions of propagation-invariant localized pulsed waves were
theoretically discovered in the late 1980-ies, and since then a massive
literature has been devoted to them. (See collective monographs \cite{LWI,LW2}
and reviews
\cite{DonelliSirged,revPIER,revSalo,MeieLorTr,KiselevYlevde,AbourClassif2019}%
). The realizability of them in optics was first demonstrated in
Ref.~\cite{PRLmeie} for the so-called Bessel-X pulse which is the only
propagation-invariant pulsed version of the monochromatic Bessel beam
introduced in Ref.~ \cite{Durnin}. The group velocity of Bessel-X pulses
exceeds $c$, i.e., it is superluminal in empty space without\ the presence of
any resonant medium. This strange property has been widely discussed in the
literature referred to above and was experimentally verified by several groups
\cite{exp2,exp3,meieXfemto,meieOPNis} for cylindrically symmetric 3D pulses.
For 2D (light sheet) counterparts of such propagation-invariant pulses
measurements of various group velocities have been recently carried out by
Abouraddy's group at the University of Central Florida
\cite{Xsheet,KondakciArbitV,AbourVisC2019}.

In our recent paper \cite{POY2019}, the following question was dealt with: how
is the group velocity of propagation-invariant pulses related---if it is
related at all---to the energy flow velocity in them? Definitely, the
statement \textquotedblleft if an energy density is associated with the
magnitude of the wave ... the transport of energy occurs with the group
velocity, since that is the rate of which the pulse travels
along\textquotedblright\ (citation from Ref.~\cite{Jackson}, section 7.8)
cannot hold if the group velocity exceeds $c$. For, no electromagnetic field
can transport energy faster than $c$ even in the case of superluminal pulses
\cite{Lekner2002,Yannis2008,YannisLWII} (see also Ref.~\cite{POY2019}).

Indeed, it is shown in Ref.~\cite{POY2019} that the velocity $v_{e}$, with
which energy in non-diffracting pulsed EM and scalar waves locally flows in
the direction of propagation, is not equal to the propagation velocity (group
velocity $v_{g}$) of the pulse itself, but these two quantities obey the
simple relation%

\begin{equation}
\frac{v_{e}}{c}=\frac{2\beta}{1+\beta^{2}}\text{ ,} \label{vevsvg}%
\end{equation}
where $\beta=v_{g}/c$. Here, we summarize the properties of this physically
content-rich relation in Fig.~1. We see that regardless of whether the pulse
propagates subluminally $(\beta<1)$ or superluminally $(\beta>1),$the energy
flow velocity $v_{e}$ does not exceed $c$. But it is surprising that the value
of $v_{e}$ is the same for both a subluminal and a superluminal pulse: this is
seen from the mirror symmetry of the curve $v_{e}/c$ with respect to the
vertical line at $v_{e}/c=1$ of the log-log plot in Fig.~1 and results from
the invariance of Eq.~(\ref{vevsvg}) with respect to the replacement
$\beta\rightarrow1/\beta$.

\begin{figure}[h]
\centering
\includegraphics[width=6cm]{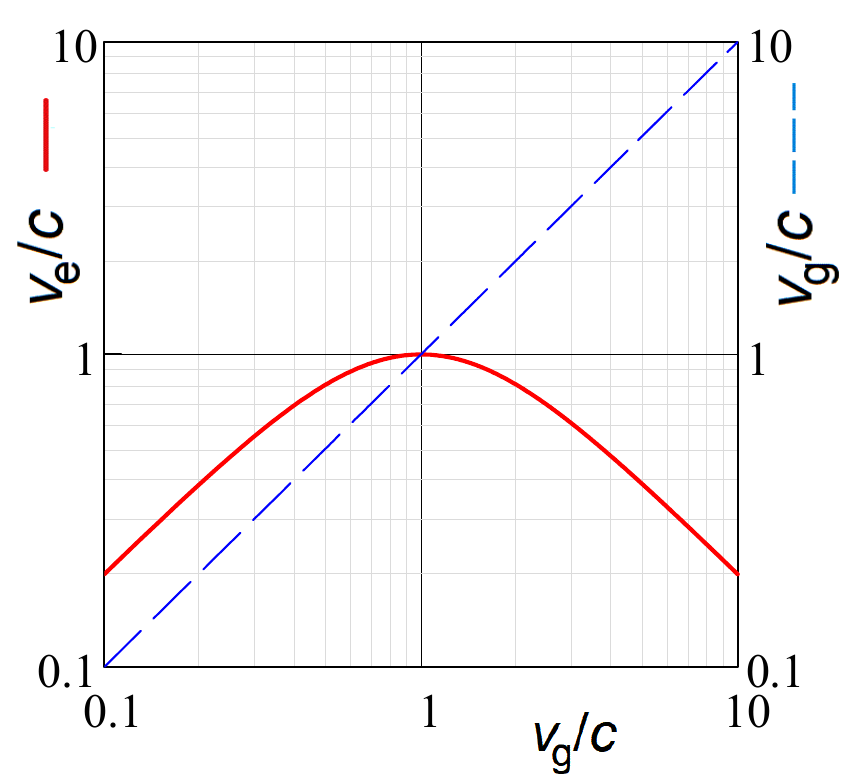}\caption{Plot of Eq. (\ref{vevsvg}) in
decimal logarithmic scales. The curve of the energy flow velocity is
juxtaposed to the straight line of unit slope, which would represent the
energy flow velocity versus the group velocity if they were equal.}%
\end{figure}

It follows from Fig.~1 that in the case of a superluminal pulse energy
propagates slower than the pulse itself and in the case of a subluminal pulse
energy propagates faster than the pulse itself. Such counterintuitive
situations as if energy is lagging behind a superluminal pulse, or as if
outrunning a subluminal pulse, are explained in Ref.~\cite{POY2019} by the
presence of the reactive energy in nondiffracting fields.

Reactive energy near sources of electromagnetic fields and in free standing
waves was thoroughly studied by Kaiser \cite{Kaiser2011}, who also proposed to
use the term 'energy flow velocity' for distinction from the time-independent
'energy transport velocity' which involves only averages over one period in
the time-harmonic case. He also introduced the term 'electromagnetic inertia
density' defined as the reactive energy density divided by $c^{2}$. Reactive
energy appears everywhere where there is an interference between plane wave
constituents in the free field. Such reactive energy is inherent to structured
light fields and can be interpreted as non-zero mass of photons
\cite{MinuPhotonMass,Minupeatykk,VintsLPh2015,FedorovLPh2017,VintsLPh2019}.
Kaiser's electromagnetic inertia density is in other words the density of mass
of a classical EM field, which is a Lorentz-invariant quantity
\cite{VintsLPh2015,FedorovLPh2017}.

The aim of the present paper is to study the formation and spatiotemporal
distribution of the reactive energy in typical nondiffracting pulses. We hope
that such study will enable better understanding of the counterintuitive
difference between velocities of pulse propagation and energy flow.

The paper has been organized as follows. In Section II we derive an expression
for the reactive energy density of a single-cycle propagation-invariant
transverse magnetic (TM) two-dimensional (2D) superluminal field. We start
with the 2D case, i.e., with a light sheet, not only for the simplicity of
such a model but also having in mind that pulsed light sheets have practical
value, e.g., in microscopy, and are presently studied intensively
\cite{Xsheet,KondakciSSelfH, AboudarXiv2019,AbouRefraOL2019}. Section III
deals with the reactive energy density of the X wave---the scalar-valued
version of which is the most studied superluminal localized wave. In Section
IV we consider the reactive energy density in some simple subluminal
propagation-invariant pulses. Finally, we discuss the results and, in
particular, the formation and Lorentz invariance of the reactive energy
density ---or, equivalently, of the invariant mass density---in structured
light fields. Appendices A and B consider the Lorentz invariance of the
reactive energy density for vector-valued and scalar electromagnetic fields,
respectively. The third Appendix describes the derivation of TM fields from
scalar ones.

\section{Reactive energy in a single-cycle superluminal light sheet}

We start with fields that do not depend on one lateral, say $y$, coordinate.
Although such 2D fields are simpler, their properties, relevant for our study,
are the same as those of cylindrical 3D ones.

Consider a TEM pulsed 2D wave propagating along the positive $z$ direction in vacuum,%

\begin{equation}
\mathbf{E}(x,z,t)=U(z-ct)\mathbf{e}_{x},~\mathbf{B}(x,z,t)=c^{-1}%
U(z-ct)\mathbf{e}_{y}, \label{EjaB}%
\end{equation}
where SI units are assumed, with $\varepsilon_{0}\mu_{0}=1/c^{2}$, and
$\mathbf{e}_{x}$ and $\mathbf{e}_{y}$ are unit vectors of a right-handed
rectangular coordinate system. $U$ is generally an arbitrary real localized
function of \textit{one} argument.

The energy flux density (Poynting vector) and the energy density of the field
are generally expressed in SI units by%
\begin{align}
\mathbf{S}  &  =c^{2}\varepsilon_{0}\ \mathbf{E\times B}\ ,\label{Poynting}\\
w  &  =\frac{1}{2}\varepsilon_{0}\mathbf{E}^{2}+\frac{1}{2}\varepsilon
_{0}c^{2}\mathbf{B}^{2}\ . \label{Energia}%
\end{align}
For the given simple field the expressions reduce to $\mathbf{S}%
=c\varepsilon_{0}U^{2}(z-ct)\mathbf{e}_{z}$ and $w=\varepsilon_{0}U^{2}%
(z-ct)$, respectively.

Following Refs. \cite{Kaiser2011,VintsLPh2015}, the reactive (rest) energy
density is defined as
\begin{equation}
R=\sqrt{w^{2}-\mathbf{S}^{2}/c^{2}}~, \label{Renergy}%
\end{equation}
which by using Eqs.~(\ref{EjaB}), (\ref{Poynting}), and (\ref{Energia}) turns
to zero in the given case. Thus, a plane wave pulse does not possess reactive
energy and, accordingly, the energy flow velocity is $c$---well-known results.

Let us take now a symmetrical pair of plane waves---the propagating direction
of the first one lies on the $(x,z)$ plane and is inclined by angle $+\theta$
with respect to the $z$-axis, and the second one by angle $-\theta$ on the
same plane. In this case, the coordinate $z$ in Eq.~(\ref{EjaB}) is replaced
by $z\cos\theta+/-x\sin\theta$ for a member of the pair, respectively. The
components of the vectors $\mathbf{E}$ and $\mathbf{B}$ for both waves
transform also according to the rules of rotation around the axis $y$. For the
polarizations given in Eq.~(\ref{EjaB}), the magnetic field remains polarized
along the $y$ axis. However, the electric field has both $x$ and $z$
components. Thus, we are dealing with a transverse magnetic (TM) field. The
resulting expressions for $\mathbf{S}$, $w$, and $R$ are rather cumbersome and
omitted here; for $\mathbf{S}$ and $w$ at points along $z$ axis they are given
in Ref.~\cite{POY2019}.

Instead, we have evaluated Eq.~(\ref{Renergy}) numerically, specifying the
function $U$---for clarity of the following interpretation---as an optical
single-cycle pulse of the form $U(\zeta)=A\sin\zeta,$ if $\left\vert
\zeta\right\vert \leq\pi$, $U(\zeta)=0$, if $\left\vert \zeta\right\vert >\pi
$, where $A$ is an amplitude constant with dimension [V/m], which is equated
to unity. The results are presented in Fig.\ 2.

\begin{figure}[h]
\centering
\includegraphics[width=7cm]{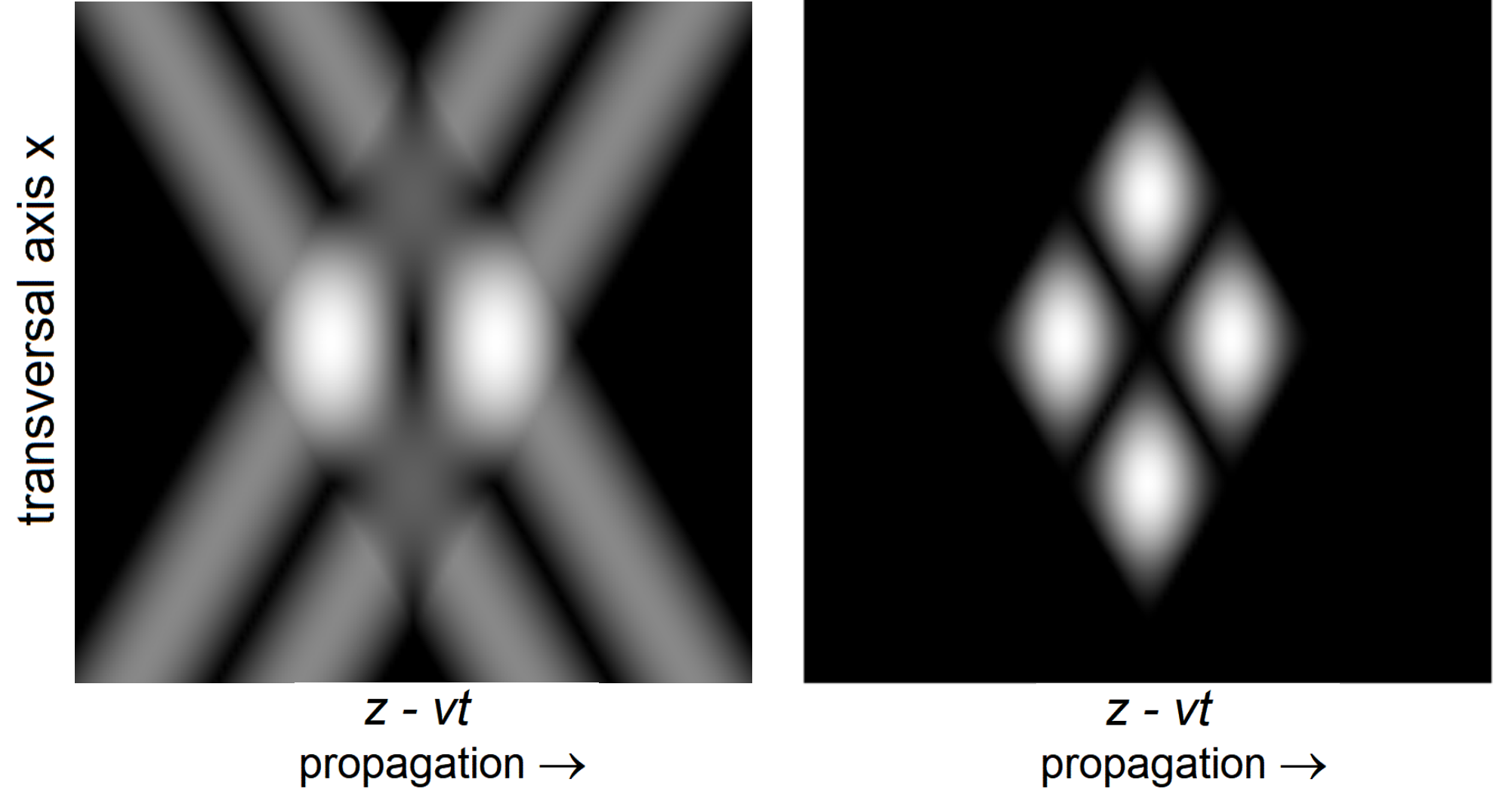}\caption{Greyscale plots of the energy
and reactive energy densities of the field of two single-cycle TM\ plane-wave
pulses propagating under angles $30^{\circ}$ and $-30^{\circ}$, respectively,
with respect to the $z$ axis. Depicted is the interference region as a
propagation-invariant superluminal 2D pulse moves along the $z$ axis with
velocity $v=c/\cos30^{\circ}=1.15c$. \ Left plot: energy density from
Eq.~(\ref{Energia}) (for lower contrast of the greyscale, $\sqrt{w}$ has been
plotted). Right plot: reactive energy density according to Eq.~(\ref{Renergy}%
). Axes $z-vt$ and $x$ have the same range \ $(6\lambda/5)\left[  -1,1\right]
$, where $\lambda$ is the wavelength of the single-cycle component pulse
(which is equal to the length of the whole pulse). The ratio of the maximum
values of the energy and reactive energy densities is $\max(w)/\max(R)=7$.}%
\end{figure}

The propagation invariance of the X-like pattern in the course of propagating
along the $z$ axis manifests itself as a dependence on the combined variable
$z-vt=z-(c/\cos\theta)t$, or on its rescaled version $z\cos\theta-ct$ (from
here on we omit the subscript $g$ denoting the group velocity for brevity
$v\equiv v_{g}$). The plots can be taken as "snapshots in flight", i.e.,
spatial distributions of the densities at a fixed time instant (say, $t=0$),
or, as temporal dependence along the transverse axis $x$ at a fixed value of
the coordinate $z$ (say, $z=0$). For clarity, below we will use the former
interpretation. Of course, we could say that the plots show purely the spatial
dependence in terms of the "co-propagating" variable $\zeta=z-vt$ and the
variable $x.$ However, such an interpretation would imply a Galilean
transformation into an unphysical superluminally moving frame, which should
not be confused with the Lorentzian transformation considered below. The
energy density at the two interference maxima is 3.5 times higher than outside
of the interference region (where it has taken to be equal to one). The ratio
would be equal to 4 if the electric field vectors of the two pulses were
parallel as the magnetic field vectors do. For the same reason, the energy
density in the two regions of destructive interference of the magnetic fields
is not zero, but has a value equal to 0.5. Understandably these ratios depend
on the value of the angle $\theta$ as will be reasoned below. Let us notice
that defining the model single-cycle pulse with abrupt beginning and end
enables one to distinctly see the edges of the interference region in the plot.

The plot of the reactive energy shows four non-zero density regions of equal
shape and maximum of 0.5. Hence, in the regions of destructive interference of
the fields the densities of energy and reactive energy have equal maxima,
where---in accordance with Eq.~(\ref{RenergyVst})---the energy flow velocity
vanishes. These equalities do not depend on the angle $\theta$. On the
contrary, the ratio of the maxima in the two plots depends on $\theta$ (or on
$\beta=v/c=1/\cos\theta$) according to the following expression%
\begin{equation}
\frac{\max(w)}{\max(R)}=\frac{1+\cos^{2}\theta}{1-\cos^{2}\theta}=\frac
{\beta^{2}+1}{\beta^{2}-1}~, \label{mw/mR}%
\end{equation}
which gives the value $7$ for $\theta=30^{\circ}$. The observations listed
above and Eq.~(\ref{mw/mR}) follow from the behavior of the field vectors
$\mathbf{E}$ and $\mathbf{B}$ in the interference region and, in particular,
from their orthogonality due to which the second term $I_{2}=$ $\mathbf{E\cdot
B}$ in Eq.~(\ref{EnergFst}) vanishes. Consequently, the reactive energy
density becomes $R=\left(  \varepsilon_{0}/2\right)  \left\vert \left(
\mathbf{E}^{2}-c^{2}\mathbf{B}^{2}\right)  \right\vert $, while the energy
density is the \textit{sum} of the two terms, see Eq.~(\ref{Energia}).
Although this holds for all subsequent 3D TM electromagnetic localized waves,
Eq.~(\ref{mw/mR}) is not valid because the maxima values of the energy and
energy densities may occur at different locations.

We see that, indeed, the reactive energy appears only where there is an
interference of the component plane waves. We conclude also from Fig.~2 that
the spatial density distribution of the reactive energy moves superluminally
in the $z$ direction and---in contradistinction to that of the energy
distribution---has no X-like "wings", i.e., is completely localized in both
directions. Moreover, as will be discussed in section 5, it is
Lorentz-invariant, i.e., the density is the same in all inertial reference
frames (see Appendix A).

\section{Reactive energy in superluminal X waves}

In this and the next section we will consider three-dimensional (3D) waves. We
start with a simple axisymmetric scalar wave pulse known under the name
"fundamental X-wave", first derived in \cite{Lu-X,Zio1993} and since then
studied by many authors. Its complex-valued expression in cylindrical
coordinates $(\rho,\varphi,z)$, normalized to unity at its maximum, reads%
\begin{equation}
\psi_{X}\left(  \rho,z,t\right)  =\frac{a}{\sqrt{\rho^{2}+\left[
a+i\tilde{\gamma}\left(  z-vt\right)  \right]  ^{2}}}~. \label{Xpsi}%
\end{equation}
Here, $\tilde{\gamma}\equiv1/\sqrt{(v/c)^{2}-1}$ is the superluminal version
of the Lorentz factor and $v$ is a superluminal group velocity identical to
the speed of\ propagation of the whole pulse. The positive free parameter $a$
determines the width of the unipolar Lorentzian-like temporal profile of the
pulse modulus on the $z$ axis. In a meridional plane $(x=\pm\rho,z)$ the
double-conical spatial profile of the field energy density looks like the
letter "X" and the branches of "X" are inclined with respect to $x$ axis under
the angle $\theta=\arccos\beta^{-1}=\arccos c/v$.

We deal first with the real scalar wave solution $u_{X}\left(  \rho
,z,t\right)  =\operatorname{Re}\psi_{X}\left(  \rho,z,t\right)  $ of
Eq.~(\ref{Xpsi}) in order to ensure Lorentz-invariance of the reactive energy
density (see Appendix B) which is essential for its physical interpretation
and comparison with the following cases of vector-valued EM fields. Fig.~3
juxtaposes the densities $w$ and $R$ calculated from Eq.~(\ref{B3b}) and
(\ref{B4}), respectively. These plots, as well as the rest in this Section,
have been obtained by introducing the dimensionless variables
$(X,Y,Z)=(x,y,z)/a$ and $T=ct/a$, using a normalized speed $c=1$, and choosing
$v/c=1.1$. Since energy densities of 3D waves---in contradistinction to the 2D
case---decrease with distance from the propagation axis $z$, for better
display of weaker features all the following figures show colored density and
surface plots.

\begin{figure}[h]
\centering
\includegraphics[width=8.5cm]{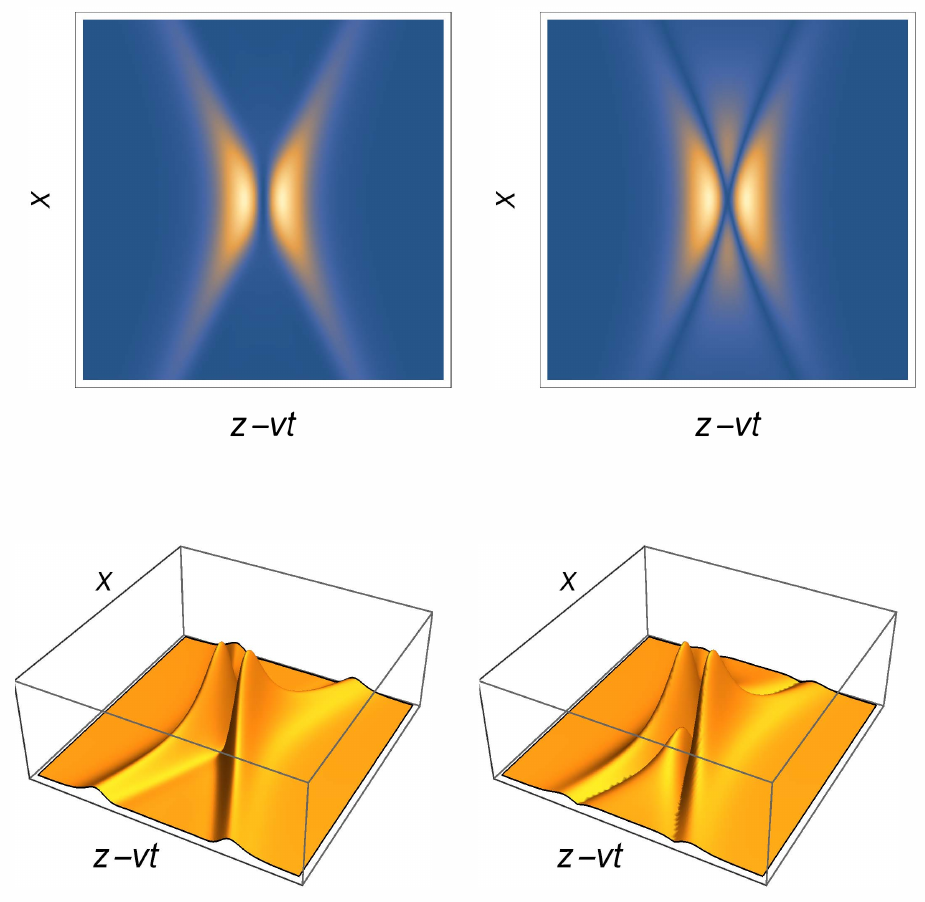}\caption{Density and surface plots of
the energy density (left) from Eq.~(\ref{B3b}) and reactive energy density
(right) from Eq.~(\ref{B4}) for the real part of the zero-order scalar X wave
for a normalized speed $c=1$ and $v/c=1.1$. Axes $z-vt$ and $x$ have the same
range \ $(2a)\left[  -1,1\right]  $ The ratio of the maximum values of the
energy and reactive energy densities is $\max(w)/\max(R)=$ $10.5$ in
accordance with Eq.~(\ref{mw/mR}).}%
\end{figure}

In the given case, since the plane wave constituents of the pulse interfere,
in principle, everywhere, the reactive energy is distributed over a large
spatial region similarly to the energy. While the reactive energy density is
generally by one order of magnitude less than that of the energy, in the
central cross-sectional plane it is not so and there we see a sharp
double-conical surface of zero reactive energy. Between the conical surfaces
$R_{\pm}$---the square of the four-gradient---becomes negative, see
Eq.~(\ref{B5}) in Appendix B.

Vector-valued X waves have been studied extensively
\cite{Recami1998,Mugnai2005,Salem2011}, but we calculated here a TM version of
the zeroth-order X wave. For comparison, Fig.~4 shows the distribution of the
energy and reactive energy densities of such a localized wave. Since obtaining
EM\ field vectors involves taking additional spatial and temporal derivatives
(see Appendix C), the region of destructive interference is now around the
apex of the cones and, consequently, the region where most of the reactive
energy is located. For this particular example, the Poynting vector and, as a
consequence the energy flow velocity, vanishes on axis ($x=0$). It follows,
then, that the energy and reactive energy densities coincide in this region.

\begin{figure}[h]
\centering\includegraphics[width=8.5cm]{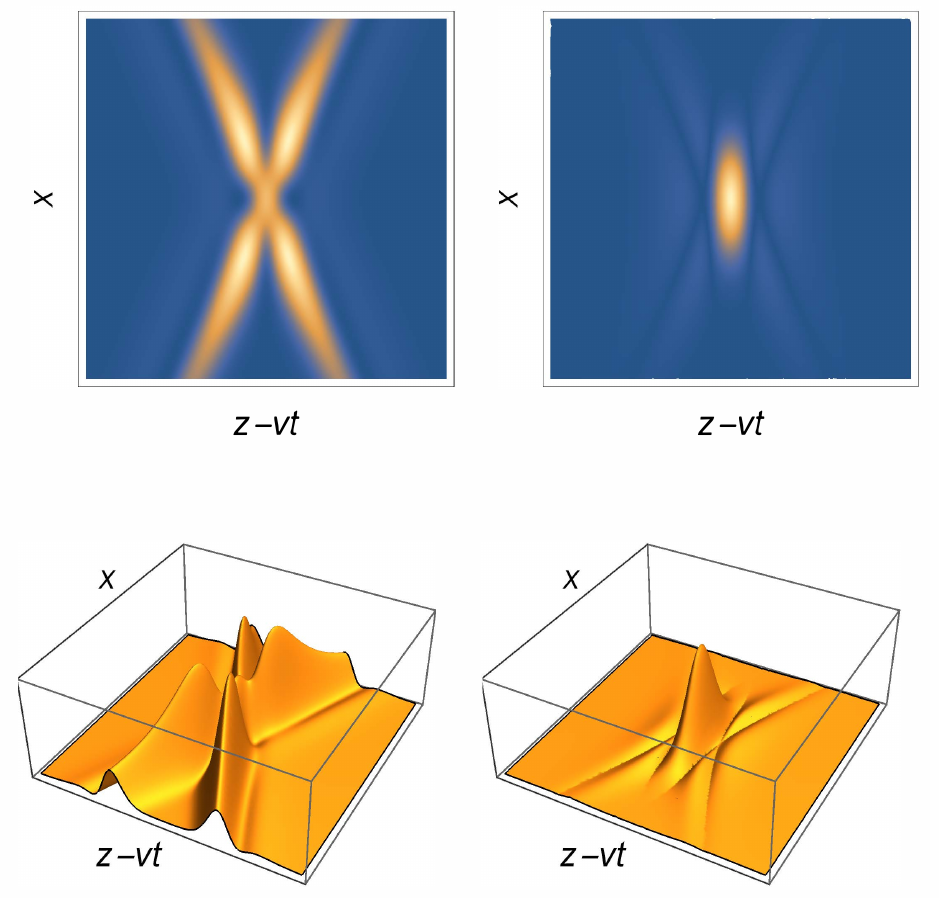}\caption{Density and surface
plots of the energy density (left) and reactive energy density (right) for a
TM zero-order vector-valued X wave for a normalized speed $c=1$ and $v/c=1.1$.
Axes $z-vt$ and $x$ have the same range\ $(2a)\left[  -1,1\right]  ~$and
$\max(w)/\max(R)=$ $1.48$. }%
\end{figure}Another physical interpretation can be given based on the
expression of the reactive energy density given in Eq.~(\ref{RenergyVst}) in
terms of the energy flow velocity. Away from the pulse center, the
spatiotemporal distribution of the modulus of the energy flow velocity is
identical to that of the energy density. In the neighborhood of the pulse
center, however, the energy flow velocity is zero, but not the energy density.
This is the reason for the concentration of the reactive energy density around
the pulse center.

For the first-order TM field, derived from an azimuthally asymmetric scalar
potential given by $\psi_{X}^{as}\left(  \rho,z,\varphi,t\right)
=(1/a)\psi_{X}\left(  \rho,z,t\right)  ^{3}\rho\exp(i\varphi)$, the
spatiotemporal distributions of the energy density and reactive energy density
are concentrated around the pulse center, albeit with different amplitudes, in
accordance with Eq.~(\ref{mw/mR}) resulting in the ratio $\max(w)/\max
(R)=10.5238$ for the energy and reactive energy densities depicted in Fig. 5
for $\varphi=0$.

\begin{figure}[h]
\centering\includegraphics[width=8.5cm]{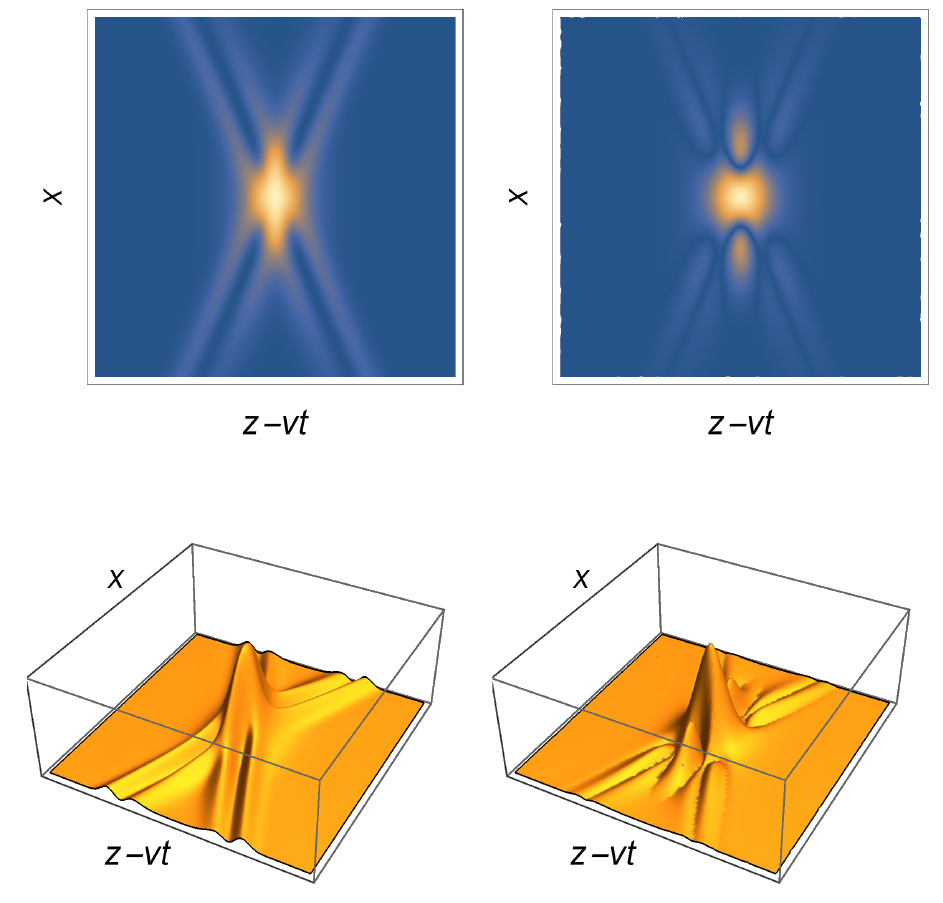}\caption{Density and surface
plots of the energy density (left) and reactive energy density (right) for a
TM first-order (asymmetric) vector-valued X wave for a normalized speed $c=1$,
$v/c=1.1$ and $\varphi=0$. Axes $z-vt$ and $x$ have the same range
\ $(2a)\left[  -1,1\right]  $ and $\max(w)/\max(R)=10.5$.}%
\end{figure}

\section{Reactive energy in subluminal waves}

The best known subluminal non-diffracting wave---the MacKinnon pulse
\cite{MacKinn}\ is a spherically symmetric monochromatic standing wave seen
from another inertial reference frame, i.e., it is obtained by applying a
Lorentz transformation with subluminal $\beta$ to the $z$-coordinate and time
\cite{revPIER,MeieLorTr}. Its real part reads
\begin{align}
\operatorname{Re}[\psi_{MK}\left(  \rho,z,t\right)  ]  &  =\frac{\sin kr_{t}%
}{kr_{t}}\cos\left[  k\beta\gamma\left(  z-\tilde{v}t\right)  \right]
,\label{McK}\\
r_{t}  &  \equiv\sqrt{\rho^{2}+\gamma^{2}\left(  z-vt\right)  ^{2}~.}\nonumber
\end{align}
Eq.~(\ref{McK}) indicates that the envelope moves rigidly along the $z$ axis
with the subluminal velocity $v=\beta c$, whereas the modulation
superluminally with $\tilde{v}\equiv\beta^{-1}c=c^{2}/v$. In addition, the
\textit{sinc}-function-like amplitude distribution is no longer spherically
symmetric as it is in the wave's rest frame, but has been compressed in the
axial direction due to the Fitzgerald-Lorentz contraction. Here, we restrict
the discussion to the scalar-valued wave. The energy and reactive energy
density plots are shown in Fig. 6. In the rest frame of the pulse, i.e., if
$v=0$, both plots would show spherical symmetry. The plots in Fig. 6 have been
obtained by introducing the dimensionless variables $(X,Y,Z)=k(x,y,z)$ and
$T=kct$, using the normalized speed $c=1$, and choosing $v/c=0.9$ and $vt=0$.

\begin{figure}[h]
\centering\includegraphics[width=8.5cm]{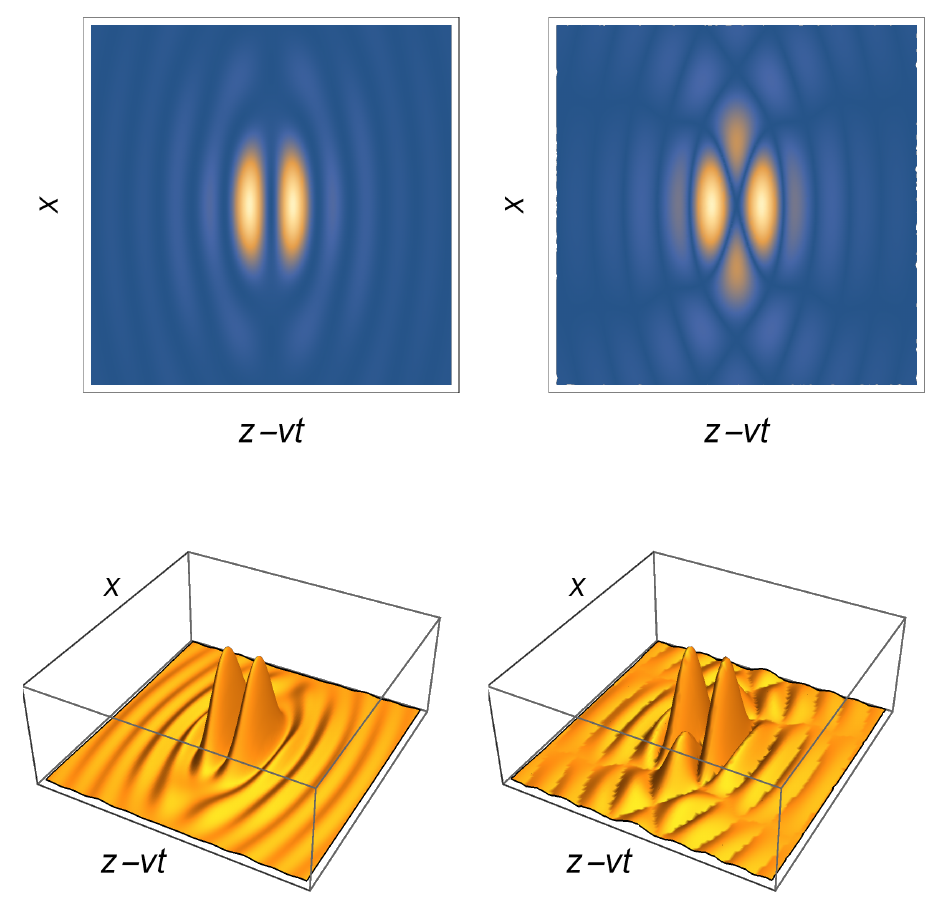}\caption{ Density and surface
plots of the energy density (left) and reactive energy density (right) for the
real part of the scalar \ MacKinnon subluminal wave packet for normalized
speed $c=1$ and $v/c=0.9$. Axes $x$ and $z-vt$ have the same range
$(5/k)[-1,1];$ $\max(w)/\max(R)=15.6$ at $t=0$.}%
\end{figure}

Due to the sinusoidal modulation of the MacKinnon pulse, the same results as
those depicted in Fig. 6 would appear periodically at $t=2\pi n\gamma/\left(
kc\right)  ,~n=1,2,3,..$. It is interesting to contrast this behavior to that
of the standing spherical wave $\sin kr\sin kct~/r$. In this case, the energy
and reactive energy densities coincide, i.e., all energy is reactive, for
$t=\pi n/\left(  kc\right)  ,~n=0,1,2,...$. It should be mentioned that
whereas the reactive energy appears predominantly in standing waves, it is
inherent also to the monochromatic expanding spherical wave $\sin\left[
k\left(  r-ct\right)  \right]  /r$ until it becomes a plane wave at infinity.

The last spatiotemporally localized field we studied was in a sense a most
physical one---the \textit{finite} energy subluminal splash mode
\cite{Splash1,Splash2,Splash3}. It arises from the elementary azimuthally
symmetric solution $(\rho^{2}+z^{2}-c^{2}t^{2})^{-1}$ of the scalar wave
equation by first resorting to the complexification $t\rightarrow t+ia$ and
subsequently undertaking a subluminal Lorentz transformation involving the
coordinates $z$ and $t$. The spatiotemporal distributions of the energy
density and the reactive energy density corresponding to the vector-valued TM
fields in this case (shown in Fig. 7) bare some similarity to those for the TM
fields of the zero-order X Wave. As in the latter case, the reactive energy
density is concentrated around the pulse center. On axis $\left(  x=0\right)
$ the Poynting vector vanishes; therefore, the energy and reactive energy
densities coincide in this region. The same similarity exists between the
scalar-valued zero-order real part of the X wave, discussed earlier, and the
scalar-valued real part of the finite-energy subluminal splash mode.

\begin{figure}[h]
\centering\includegraphics[width=8.5cm]{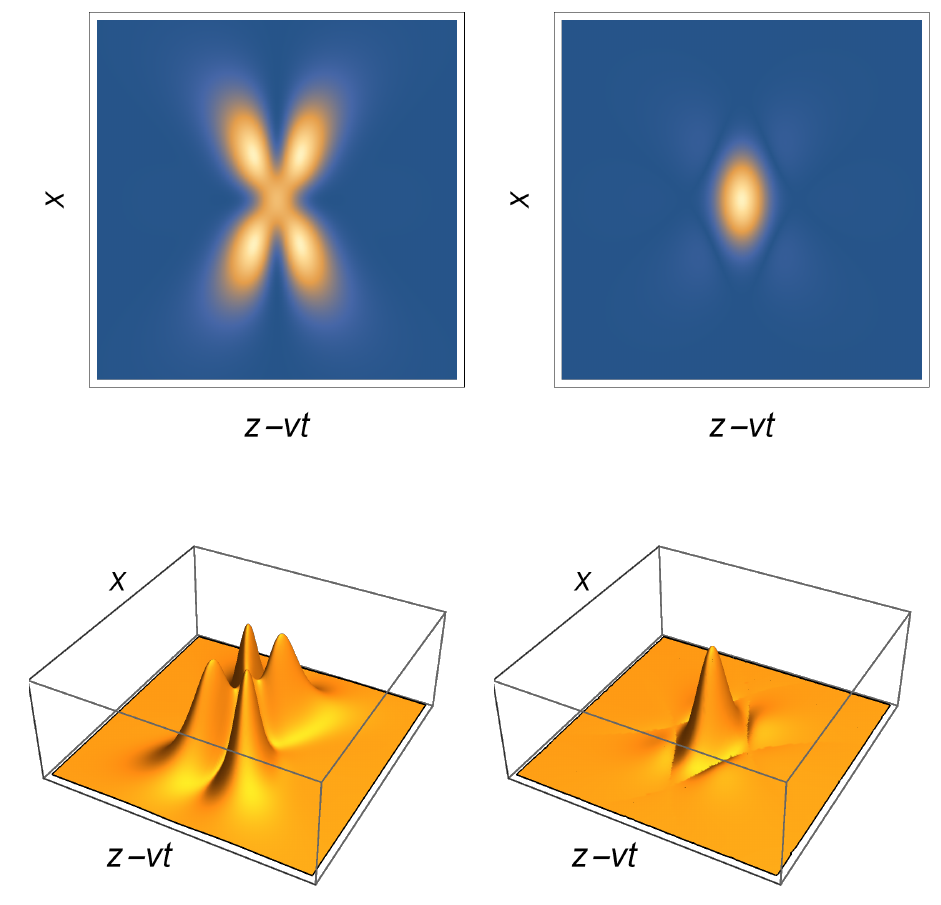}\caption{Density and surface
plots of the energy density (left) and reactive energy density (right) for a
TM finite-energy subluminal vector-valued splash mode for normalized speed
$c=1$, $v/c=0.9$ and $t=0$. Axes ranges $z-vt\in(4a/5)[-1,1]$ and
$x\in(2a)[-1,1]$; $\max(w)/\max(R)=1.43.$}%
\end{figure}

\section{Discussion}

The main and expected conclusion from all plots is that the reactive energy
density appears in the regions where interference between plane wave
constituents with different propagation directions takes place. Alternatively,
the reactive energy is minimum when the modulus of the energy flow velocity is
close to the speed of light in vacuum, and approaches the value of the energy
density in the limit as the energy flow velocity approaches zero. In a sense,
the reactive energy appears at the expense of decrease of the energy density
$w$ in the interference regions. This may serve as an explanation of the
paradoxical circumstance that the energy flow velocity is not only smaller
than the superluminal pulse propagation velocity but also smaller than $c$, as
was shown in Ref.~\cite{POY2019}.

As mentioned in Sect.~II, Fig.~2 can be equivalently considered as showing
temporal dependence of the energy densities on a plane with fixed value of
$z$, and, due to the symmetry of the plots, the time axis may be thought of as
directed also from the left to the right. For an observer in a reference frame
moving along the $z$ axis with subluminal velocity $c^{2}/v$, all 3D X-type
propagation-invariant waves are seen as first collapsing and then expanding
cylindrical pulses with $z$-\textit{independent} energy distributions
\cite{revPIER,MeieLorTr}. In the given 2D case, such an observer sees two
plane wave pulses counterpropagating (under angle $2\theta=\pi$ with respect
to each other) along the $x$ axis, and for the observer Fig.~2 shows temporal
evolution of \ distributions of both energy densities, which is independent of
the coordinate $z$. Hence, for the moving observer Fig.~2 demonstrates
explicitly that the reactive energy exists nowhere in the space initially, it
is \textit{created }when the pulses collide, and it disappears after that.
Likewise, in the laboratory frame the superluminal movement of the reactive
energy may be interpreted rather as its continuous creation and disappearance
at successive locations. This may serve as an additional argument for the
solution of the paradoxical feature of the energy flow velocity.

It should be noted that according to Eq.~(\ref{RenergyFst}) the Lorentz
invariant $I_{1}$ is negative if the magnitude of $c\mathbf{B}$ exceeds that
of the electric field $\mathbf{E}$. Thus, although by definition of the
reactive energy density through the square root it is always non-negative, in
the cases when $\mathbf{E}\cdot\mathbf{B}=0$ we may consider an alternative
definition of the density $R_{\pm}=\varepsilon_{0}\left(  \mathbf{E}^{2}%
-c^{2}\mathbf{B}^{2}\right)  /2$. For example, within the two rhombus-like
regions on the horizontal axis on the right-hand plot of Fig.~2, $R_{\pm}<0$
because the magnetic field is stronger than the electric field there. If the
pulse was unipolar, $R_{\pm}$ would be positive nowhere. The X-like cross of
dark lines in the plot that separates the four regions indicates the points
where $R=\left\vert R_{\pm}\right\vert =0$, i.e., $R_{\pm}$ changes its sign.
Similar zero lines are seen in all subsequent plots of $R$. In the case of
scalar-valued fields, as seen from Eq.~(\ref{B5}), $R_{\pm}$ becomes negative
if the magnitude of the spatial gradient exceeds the absolute value of the
time derivative of the field. Understandably, the quantity $R_{\pm}/c^{2}$
cannot be interpreted as a mass density. This fact, to a certain degree,
undermines the known interpretation \cite{Kaiser2011,VintsLPh2015} of
$R/c^{2}$ as the rest mass density of the field.

In the case of infinite-energy non-diffracting fields, it does not make much
sense to ask what is the total reactive energy. However, all such fields
become finite-energy ones in reality due to the finite apertures of the
devices generating them. A finite aperture ensures a finite depth and a finite
duration of diffraction-free propagation. Outside of the region of
diffraction-free propagation the pulse develops into a spherical wave with
velocity $c$ which becomes a plane wave at infinity. Thus, with reference to
Eq.~(\ref{RenergyVst}), we can conclude that the spatio-temporal region where
the reactive energy is not vanishingly small, is always finite for real
pulses. Despite the fact that the reactive energy density is
Lorentz-invariant, the total reactive energy $\int Rdxdydz$ is not
\cite{FedorovLPh2017}. However, following the example of the invariant action
as four-integral of the Lagrangian density, we can construct the quantity
$\int R_{\pm}dxdydzdt$ which is Lorentz-invariant and might be called
'reactive action'. It is interesting to note that if $\mathbf{E\cdot B=0}$,
the reactive energy density $R_{\pm}$ coincides with the Lagrangian density of
the field and, hence, the 'reactive action' is equal to the action. On the
other hand, it is known that if one \textit{first} integrates over all space
separately the energy and momentum density of the pulse, and then takes the
difference of their squares, i.e., the difference of squares of total energy
$W$ and total momentum multiplied by $c$, \textit{viz.}%
\begin{equation}
\left(  \int wdxdydz\right)  ^{2}-\left(  \frac{1}{c}\int\mathbf{S}%
dxdydz\right)  ^{2}=W^{2}-(c\mathbf{P})^{2}~, \label{GlobInvt}%
\end{equation}
one gets a quantity which is the same in all inertial reference frames (see,
e.g., Ref.~\cite{LekneriRmt}). The quantity given in Eq.~(\ref{GlobInvt}) is
the square of the total reactive energy and its division by $c^{4}$ can be
considered as the square of the invariant rest mass of the pulse
\cite{Kaiser2011,VintsLPh2015,FedorovLPh2017}. The reasons why the integral
$\left(  \int Rdxdydz\right)  ^{2}$calculated using Eq.~(\ref{Renergy}) does
not give the same invariant quantity are obvious. Mathematically, two
operations---integration and taking the difference of squares---are carried
out in different order in either case. Physically, Eq.~(\ref{GlobInvt})
characterises reactive energy globally, while Eq.~(\ref{Renergy}) ---locally.

Considering, finally, the complex time-dependent Poynting vector as
$\mathbf{S}\left(  \mathbf{r},t\right)  =\mathbf{E}\left(  \mathbf{r}%
,t\right)  \times\mathbf{H}^{\ast}\left(  \mathbf{r},t\right)  $, we obtain
the \textquotedblleft conservation\textquotedblright\ law%
\begin{equation}
\nabla\cdot\mathbf{S}+\varepsilon_{0}\left(  \mathbf{E\cdot}\frac{\partial
}{\partial t}\mathbf{E}^{\ast}+c^{2}\mathbf{B\cdot}\frac{\partial}{\partial
t}\mathbf{B}^{\ast}\right)  =0~. \label{CLawcomp}%
\end{equation}
Writing the complex Poynting vector explicitly in terms of its real and
imaginary parts, viz., $\mathbf{S=S}_{r}+i\mathbf{S}_{i}$, we obtain%
\begin{equation}
\nabla\cdot\mathbf{S}_{r}+\frac{\partial}{\partial t}w_{r};\quad w_{r}%
=\frac{\varepsilon_{0}}{2}\left(  \mathbf{E\cdot E}^{\ast}+c^{2}\mathbf{B\cdot
B}^{\ast}\right)  ~, \label{CLawRe}%
\end{equation}
and for the imaginary part of the Poynting vector%
\begin{equation}
\nabla\cdot\mathbf{S}_{i}+\varepsilon_{0}\operatorname{Im}\left\{
\mathbf{E\cdot}\frac{\partial}{\partial t}\mathbf{E}^{\ast}+c^{2}%
\mathbf{B\cdot}\frac{\partial}{\partial t}\mathbf{B}^{\ast}\right\}  =0~.
\label{CLawIm}%
\end{equation}
For time-harmonic fields, Eq.~(\ref{CLawIm}) would result in the conventional
reactive energy density.

\section{Conclusions}

The notion of \textquotedblleft reactive energy\textquotedblright\ has been of
fundamental importance in assessing the efficiency of monochromatic radiating
structures by evaluating the stored (non-propagating) energy in the near field
\cite{Capek2016}. With the widespread applications of wideband pulsed signals,
attempts have been made recently to expand the concept of reactive energy to
the time domain. Different approaches have been followed along this direction.
Recently Mikki, Sarkar and Antar \cite{Mikki2019}, for example, have
introduced the \textquotedblleft localized energy density\textquotedblright%
\ defined as the difference of the electromagnetic energy density and the
modulus of the Poynting vector divided by $c$ in free space.

In this article, we have adopted the definition of the reactive energy density
$R=\sqrt{w^{2}-\mathbf{S}^{2}/c^{2}}$, see Eqs.~(\ref{Poynting}%
)-(\ref{Renergy}), introduced by Kaiser \cite{Kaiser2011} by analogy to the
rest energy $E_{0}=\sqrt{E^{2}-c^{2}\mathbf{p}^{2}}$ of a relativistic point
particle having total energy $E$ and momentum $\mathbf{p}$. The ratio
$R/c^{2}$ is defined as the rest mass density of the fields
\cite{Kaiser2011,VintsLPh2015}. The reactive energy density is Lorentz
invariant and for fields where $\mathbf{E\cdot B=0}$, the reactive energy
density coincides with the Lagrangian density. Proofs of the Lorentz
invariance are given in Appendices A and B for vector-valued and scalar-valued
electromagnetic fields, respectively.

The definition of the time-dependent reactive energy density $R$ in
Eq.~(\ref{Renergy}) has been used in this article mostly as a measure of
\textquotedblleft interference\textquotedblright\ in order to explain and
illustrate the difference between the energy flow velocity and the group
velocity in structured light pulses. Specific numerical results have been
presented for basic types of scalar-valued and vector-valued superluminal and
subluminal spatiotemporally localized electromagnetic waves in vacuum. The
main conclusion has been that the reactive energy density is larger in regions
of strong interference of the constituent plane wave components of the
structured light field pulses. Alternatively, the reactive energy is minimum
when the modulus of the energy flow velocity is close to the speed of light in
vacuum and approaches the value of the energy density in the limit as the
energy flow velocity approaches zero.

\appendix

\section{PROOF OF LORENTZ INVARIANCE OF REACTIVE ENERGY DENSITY FOR EM FIELDS}

Reactive energy and its quotient by $c^{2}$---reactive mass density---are the
same in different inertial reference frames \cite{Kaiser2011,VintsLPh2015}.
Here we give another proof of this remarkable property and relate these
quantities to energy flow and group velocities.

Let us compose the complex-valued Riemann-Silberstein vector from the
real-valued electric and magnetic fields satisfying the homogeneous Maxwell
equations in free space%
\begin{equation}
\mathbf{F}=\sqrt{\frac{\varepsilon_{0}}{2}}\left(  \mathbf{E}+ic\mathbf{B}%
\right)  ~. \label{RSvect}%
\end{equation}
Certain important physical quantities associated with the real fields
$\mathbf{E}\left(  \mathbf{r},t\right)  $ and $\mathbf{B}\left(
\mathbf{r},t\right)  $ can be expressed conveniently in terms of
$\mathbf{F}\left(  \mathbf{r},t\right)  $. Specifically, the Poynting vector
$\mathbf{S}$ and the electromagnetic field energy density $w$---see also
Eqs.~(\ref{Poynting}) and (\ref{Energia})--- can be written as
\begin{align}
\mathbf{S}  &  =-ic\mathbf{F}^{\ast}\times\mathbf{F~,}\nonumber\\
w  &  =\mathbf{F}^{\ast}\cdot\mathbf{F~,} \label{EnergFst}%
\end{align}
where the asterisk denotes complex conjugation. The reactive energy density
given in Eq.~(\ref{Renergy}) can be written in terms of the
Riemann-Silberstein vector and the fields as%
\begin{align}
R  &  =\left\vert \mathbf{F}\cdot\mathbf{F}\right\vert =\sqrt{\frac
{\varepsilon_{0}^{2}}{4}I_{1}^{2}+\varepsilon_{0}^{2}c^{2}I_{2}^{2}%
}~;\label{RenergyFst}\\
I_{1}  &  =\mathbf{E}^{2}-c^{2}\mathbf{B}^{2},\quad I_{2}=\mathbf{E\cdot
B~.}\nonumber
\end{align}
Both $I_{1}$ and $I_{2}$ are well-known Lorentz invariants. As consequence,
the reactive energy density $R$ is Lorentz-invariant. It can also be expressed
as%
\begin{equation}
R=w\sqrt{1-\frac{\left\vert \mathbf{v}_{e}\right\vert ^{2}}{c^{2}}}~,
\label{RenergyVst}%
\end{equation}
in terms of the electromagnetic energy density and the local energy flow
velocity $\mathbf{v}_{e}$. In locations where $\mathbf{v}_{e}$ is parallel to
$z$ axis and Eq.~(\ref{vevsvg}) holds, we obtain%
\begin{equation}
R=w\left\vert \frac{1-\beta^{2}}{1+\beta^{2}}\right\vert ~,
\label{Renergybeetast}%
\end{equation}
where $\beta=v_{g}/c$ is the normalized group velocity. It follows, then, that
the reactive energy density is maximum (equal to $w$) when $\mathbf{v}_{e}=0$
(or $v_{g}=0$, i.e., the pulse is not moving in the $z$ direction) and equals
zero when $\mathbf{v}_{e}=c$ (or $v_{g}=c$). The latter is the case for null
electromagnetic waves, which are also characterized by the fact that both
$I_{1}$and $I_{2}$ equal zero.

\section{PROOF OF LORENTZ INVARIANCE OF REACTIVE ENERGY DENSITY FOR SCALAR
FIELDS}

Optical fields, especially paraxial ones, can be in good approximation
described by a single scalar function $\psi(x,y,z,t)$. In this case, the
Poynting vector $\mathbf{S}$ and energy density $w$ are given by the
expressions \cite{GrnW1953,MW}.%

\begin{subequations}
\begin{align}
\frac{\mathbf{S}}{\alpha c}  &  =-\left(  \partial_{ct}\psi^{\ast}\right)
\left(  \nabla\psi\right)  -\left(  \partial_{ct}\psi\right)  \left(
\nabla\psi^{\ast}\right)  ~,\label{B1a}\\
\frac{w}{\alpha}  &  =\left(  \partial_{ct}\psi\right)  \left(  \partial
_{ct}\psi^{\ast}\right)  +\left(  \nabla\psi\right)  \cdot\left(  \nabla
\psi^{\ast}\right)  ~, \label{B1b}%
\end{align}
where $\partial_{ct}$ denotes the derivative with respect to $ct$, and
$\alpha$ is a positive constant whose value depends on the choice of units.
The expressions have been written so that the right-hand sides are of the same dimension.

Using Eqs.~(\ref{Renergy}), (\ref{B1a}), and (\ref{B1b}), we obtain for square
of the reactive energy density the following expression:%

\end{subequations}
\begin{align}
\frac{R^{2}}{\alpha^{2}}  &  =\left[  \left(  \partial_{ct}\psi\right)
^{2}-\left(  \partial_{x}\psi\right)  ^{2}-\left(  \partial_{y}\psi\right)
^{2}-\left(  \partial_{z}\psi\right)  ^{2}\right] \nonumber\\
&  \times\left[  \left(  \partial_{ct}\psi^{\ast}\right)  ^{2}-\left(
\partial_{x}\psi^{\ast}\right)  ^{2}-\left(  \partial_{y}\psi^{\ast}\right)
^{2}-\left(  \partial_{z}\psi^{\ast}\right)  ^{2}\right] \nonumber\\
&  -\left[  \left(  \partial_{x}\psi^{\ast}\right)  \left(  \partial_{y}%
\psi\right)  -\left(  \partial_{y}\psi^{\ast}\right)  \left(  \partial_{x}%
\psi\right)  \right]  ^{2}\label{B2}\\
&  -\left[  \left(  \partial_{x}\psi^{\ast}\right)  \left(  \partial_{z}%
\psi\right)  -\left(  \partial_{z}\psi^{\ast}\right)  \left(  \partial_{x}%
\psi\right)  \right]  ^{2}\nonumber\\
&  -\left[  \left(  \partial_{z}\psi^{\ast}\right)  \left(  \partial_{y}%
\psi\right)  -\left(  \partial_{y}\psi^{\ast}\right)  \left(  \partial_{z}%
\psi\right)  \right]  ^{2}~.\nonumber
\end{align}
The first term on the right-hand side is Lorentz invariant, but not the
remaining three terms. Therefore, the reactive energy density of the
complex-valued scalar field $\psi$ is not Lorentz invariant. It should be
noted, however, that for a real field $u(x,y,z,t)$ which can be either the
real or the imaginary part of $\psi(x,y,z,t)$, the Poynting vector and energy
density are defined as
\begin{align}
\frac{\mathbf{S}}{\alpha c}  &  =-2\left(  \partial_{ct}u\right)  \left(
\nabla u\right)  ~,\label{B3a}\\
\frac{w}{\alpha}  &  =\left(  \partial_{ct}u\right)  ^{2}+\left(  \nabla
u\right)  \cdot\left(  \nabla u\right)  ~. \label{B3b}%
\end{align}
In this case, the reactive energy density $R$ is Lorentz invariant because
with taking $\alpha=1$, Eq.~(\ref{B2}) reduces to
\begin{equation}
R=\left\vert \left(  \partial_{ct}u\right)  ^{2}-\left(  \partial_{x}u\right)
^{2}-\left(  \partial_{y}u\right)  ^{2}-\left(  \partial_{z}u\right)
^{2}\right\vert ~. \label{B4}%
\end{equation}
It is interesting to note that in the case of scalar fields, as it follows
from Eq.~(B2), one can define a quantity%
\begin{equation}
R_{\pm}=\left(  \partial_{ct}u\right)  ^{2}-\left(  \partial_{x}u\right)
^{2}-\left(  \partial_{y}u\right)  ^{2}-\left(  \partial_{z}u\right)  ^{2}~,
\label{B5}%
\end{equation}
which is also Lorentz invariant with dimension of energy, but can acquire
negative values in regions where the square of the four-gradient is negative.
Let us add that although the quantity $\left\vert \psi(x,y,z,t)\right\vert
^{2}$ is often used in the role of the energy density, it is correct only for
paraxial scalar fields.

To our best knowledge, the results of this Appendix have been obtained for the
first time.

\section{TRANSVERSE MAGNETIC (TM) VECTOR FIELDS}

A real-valued electric vector Hertz potential $\mathbf{\Pi}_{e}(\mathbf{r},t)$
is defined as follows:%
\[
\mathbf{\Pi}_{e}(\mathbf{r},t)=\mathbf{e}_{z}\operatorname{Re}\left\{
\psi(\mathbf{r},t)\right\}  .
\]
Here, $\mathbf{e}_{z}$ is a unit vector along the $z$ direction and
$\psi(\mathbf{r},t)$\ is a subluminal or superluminal localized solution to
the scalar wave equation in free space. The corresponding real TM
electromagnetic fields are given by
\begin{align*}
\mathbf{E}(\mathbf{r},t)  &  =\nabla\nabla\cdot\mathbf{\Pi}_{e}(\mathbf{r}%
,t)-\frac{1}{c^{2}}\frac{\partial^{2}}{\partial t^{2}}\mathbf{\Pi}%
_{e}(\mathbf{r},t)\ ,\\
\mathbf{H}(\mathbf{r},t)  &  =\varepsilon_{0}\nabla\times\frac{\partial
}{\partial t}\mathbf{\Pi}_{e}(\mathbf{r},t)\ .
\end{align*}

With $\mathbf{B}(\mathbf{r},t)=\mu_{0}\mathbf{H}(\mathbf{r},t)$\ in vacuum,
the energy and reactive energy densities can be computed using
Eqs.~(\ref{RSvect})-(\ref{RenergyFst}). Due to the invariance of the energy
flux and energy density with respect to the duality transformation
$\mathbf{E}\rightarrow c\mathbf{B}$, $\mathbf{B}\rightarrow-\mathbf{E/}c$ the
results obtained in this paper for TM pulses apply also for TE pulses.

\end{document}